\documentstyle[12pt]{article} \headheight 0cm \headsep 0cm
\newlength{\mytopmargin} \newlength{\myleftmargin}
\begin{document}  \rightline{UBCTP-93-25}
\noindent {\large {\bf Conformal Field Theory Approach to Quantum Impurity
Problems$^*$ } \vspace{5.5mm}}

\normalsize \noindent Ian Affleck  \vspace{5mm}

\noindent Canadian Institute for Advanced Research and Physics Department,
University of British Columbia, Vancouver, B.C., V6T1Z1, Canada
 \vspace{25.5mm}

\normalsize \rm \noindent
   A brief review is given of a new method for studying the critical
behavior of quantum impurity problems, based on conformal field theory
techniques, which I developed with Andreas Ludwig.  Some results on the
overscreened Kondo problem are reviewed.  It is shown that the simple open
and periodic fixed points, which occur in quantum spin chain impurity
models, are related to each other by fusion.

 \vspace{10mm} \noindent {\bf 1. Introduction } \vspace{5mm}

\noindent Quantum impurity problems occur in various areas of physics.  Some
examples are: an impurity spin in a metal (the Kondo problem), the
two-impurity Kondo problem, a defect or impurity spin in a spin chain,
tunnelling through a barrier in a quantum wire proton-monopole scattering
(the Callan-Rubakov effect).  Some common features of these problems are:
\newline 1.  gapless ``bulk'' excitations (ie. far from the impurity or in
the absence of an impurity the excitations are gapless),
 \newline 2. essentially harmonic behavior away from the impurity (Fermi or
Luttinger liquids), \newline 3. the impurity is localized and may carry
quantum-mechanical degrees of freedom (eg. the two spin states of the
impurity spin), \newline 4.perturbation theory in the impurity-bulk coupling
may be infrared divergent, \newline 5. the problems are either initially
defined in one dimension or can be mapped into one dimension (ie. some sort
of partial wave expansion can be made and only a finite number of partial
waves are important).

The first and last points imply that the systems are equivalent to
$(1+1)$-dimensional conformal field theories with the impurity or defect at
the origin of the one-dimensional position space.  In the imaginary time
path-integral formulation, we have a boundary or defect line at $(\tau ,0)$.
See Figure 1.  John Cardy recently developed [1] a boundary conformal field
theory to treat two-dimensional classical critical systems with boundaries.
We have extended his technique to deal with the $(1+1)$ dimensional quantum
case, allowing for the possibility of a dynamical defect [2-14].

In the next section we review a simple and venerable example to provide some
motivation: the local Fermi liquid theory of the Kondo effect, developed by
Nozi\`eres [15], following earlier ideas of Wilson and Anderson.  In Section
3 we discuss our general approach to quantum impurity problems.  In Section
4 we present some of our results on the overscreened Kondo effect [2-6,\
8-10], in order to demonstrate the power of the new method.  In Section 5 we
discuss impurities in the spin $s=1/2$ Heisenberg antiferromagnetic chain
[11,\ 12].  In particular we show that the simple open and periodic fixed
points which occur in the problem are related by the ``fusion'' technique,
which we used in our analysis of the Kondo problem.

\vspace{10mm}
 \noindent {\bf 2. A Simple Example: Local Fermi Liquid Theory of the Kondo
Effect} \vspace{5mm}

\noindent The continuum form of the Hamiltonian density is:

$$ {\cal H} =\psi^\dagger \left( -{\nabla^2\over 2m}\right)\psi
+J\delta^3(\vec x)\psi^\dagger {\vec \sigma \over 2}\psi \cdot \vec S
\eqno(1)$$

\noindent Here $\vec S$ is an $s=1/2$ spin operator and the fermion
annihilation operator, $\psi_\alpha$ carries a spin index which is not
written explicitly. Expanding in spherical harmonics, only the $s$-wave
interacts with the impurity due to the $\delta$-function.  Thus we obtain an
effective one-dimensional model defined on the half line, $r\geq 0$ with the
impurity at the origin. Alternatively, we may reflect the outgoing wave to
the negative $r$-axis, obtaining a theory with a left-mover only on the
entire real axis.  (See Figure 2.)  The corresponding Hamiltonian density,
written in terms of a left-moving fermion operator, $\psi$ is:

$$ {\cal H}=\psi^\dagger i{d\over dx}\psi +\lambda \delta (x) \psi^\dagger
{\vec \sigma \over 2}\psi \cdot \vec S  \eqno(2)$$

\noindent As discovered by Kondo [16], perturbation theory in the Kondo
coupling, $\lambda$, is infrared divergent.  In modern language, this
corresponds to the $\beta$-function:

$${d\lambda \over d \hbox{ln} L}=\lambda^2+ ... \eqno(3)$$

\noindent It appears that $\lambda$ renormalizes to $\infty$ if it is
initially positive. (See Figure 3.)  What does an infinite effective coupling
constant really mean?  Nozi\`eres made this notion precise, [15], by
considering a lattice version of the Kondo problem.  Since the
dimensionality of the lattice is unimportant we will consider the
one-dimensional case:

$$H =
t\sum_i\left(\psi^\dagger_i\psi_{i+1}+\hbox{h.c.}\right)+J\psi^\dagger_0{\vec
\sigma \over 2}\psi_0 \cdot \vec S \eqno(4)$$

\noindent This model is easy to study for $J>>t$.  For $t=0$ we must have
precisely one electron at the origin, which forms a singlet with the impurity
spin, $\vec S$.  The electronic configuration on the other sites is
arbitrary.  For a relatively small non-zero $t$, this degeneracy is broken.
The other electrons simply go into a Bloch wave (Slater determinant) state,
but the single-particle wave-functions must all vanish at $x=0$ in order to
preserve the spin-singlet condition there.  Thus is the even-parity sector,
the zero-coupling wave function:

$$ \phi (x) = \cos kx \ \ (\lambda = 0)\eqno(5)$$

\noindent gets modified to:

$$ \phi (x) = |\sin kx| \ \ (\lambda \to \infty ).\eqno(6)$$

\noindent This is a solution of the free-particle Schroedinger equation
everywhere except at the origin where the vanishing boundary condition is
imposed. On the other hand, the odd-parity wave-function:

$$ \phi (x) = \sin kx \eqno(7)$$

\noindent is the same for zero {\it or} infinite coupling.
 As Nozi\`eres observed [15], {\it the strong coupling fixed point is the
same as the weak coupling fixed point except that the impurity disappears
(screened) and is replaced by a Boundary Condition},

$$ \psi (0) = 0. \eqno(8)$$

\noindent Equivalently, we have a $\pi /2$ phase shift in the even parity
channel.

For finite (or even small) Kondo coupling this picture still holds but only
at low energies and long distances.  The {\it boundary condition} is a {\it
fixed point}.  ie., we have an {\it effective boundary condition} which
holds in the asymptotic regime.  Our main new result is that {\it All Quantum
Impurity Problems work this way}!

 \vspace{10mm} \noindent {\bf 3. General Approach } \vspace{5mm}

\noindent Far from the impurity we expect the low energy physics to be
described by a scale-invariant, time-independent boundary condition.  This
should be true outside a boundary layer whose width, $a$, is set by the
longest microscopic or crossover length scale in the problem.  Further from
the impurity than this it is reasonable to expect scale invariance.  However
the critical behavior may still be affected, in a universal way, by the
presence of the boundary.  Consider, for example, a two-point Green's
function.  As illustrated in Figure 4  there are two special limits.  In the
limit where both points are far from the impurity relative to their distance
from each other  we expect to recover the bulk critical behavior, unaffected
by the boundary.  However, in the opposite limit where the two points are
far apart compared to their distance from the boundary (which is still large
compared to the width, $a$, of the boundary layer), the critical behavior
can be quite different.  ie. boundary critical phenomena can occur.  In
general, to each bulk universality class corresponds several boundary
universality classes.  We expect essentially arbitrary boundary interactions
should renormalize to one of these.  A simple example is provided by the
classical two dimensional Ising model, at temperature, $T=T_c$, defined on
the half-plane, $x>0$.  There are only three universality classes of
boundary conditions, spin up, spin down and free [1].   The correlation
exponent, $\eta =1/4$ far from the boundary, but $\eta =1$ in the boundary
limit, for free boundary conditions.  Furthermore, the universal cross-over
function, depending only on the ratios of distances from the boundary to
distance between the points has been calculated exactly, by Cardy [1].  If
we apply a weak magnetic field near the boundary, the system should cross
over to the spin up universality class at large distances.

For analogous reasons to the bulk case, we expect not only translational and
scale invariance but the infinite-dimensional conformal symmetry to hold at
the critical point.  We introduce a complex co-ordinate:

$$ z\equiv \tau + ix \eqno(9)$$

\noindent A general conformal transformation is of the form: $z \to w(z)$
where $w(z)$ is an analytic function.  This has the Taylor expansion:

$$w(z) = \sum_{n=1}^\infty a_nz^n \eqno(10)$$

\noindent where the $a_n$'s are arbitrary complex numbers.  With a boundary
at $x=0$ we can, at most, expect the subgroup of the conformal group which
leave the boundary invariant to be a symmetry of the problem.  ie. we
require $w(\tau )^*=w(\tau )$.  This implies that the expansion coeffients,
$a_n$ are all real.  Since the $a_n$'s correspond to generators of the
conformal group, we see that one half the conformal group remains in the
presence of the conformally invariant boundary condition.  {}From the
real-time, Hamiltonian viewpoint, the boundary relates left and right
movers.  Consequently, the left and right energy density, $T$ and $\bar T$
are no longer independently conserved.  There is now only one Virasoro
algebra, not two as in bulk conformal field theory.

More explicitly, let us assume that the problem is defined on the positive
half-plane, $x>0$ only.  (We can always represent the problem this way by
reflecting the other half-plane, if neccessary.)  We will then assume the
condition:

$$T(0,t) =  \bar T(0,t)\eqno(11)$$

\noindent This results from assuming that the momentum density vanishes at
the boundary; it is essentially a unitarity condition. Since $T$ is  a
function of $t+x$ only and $\bar T$ is a function of $t-x$ only, Eq. (11)
implies that we may regard $\bar T$ as the analytic continuation of $T$ to
the negative $x$-axis; ie.

$$T(t,x) \equiv \bar T(t,-x) \ \ \hbox{for}\ x<0\eqno(12)$$

\noindent Thus we obtain a problem defined on the entire real axis with
left-movers only.  This identification of left with right leads to a
modification of Green's functions near the boundary.  An arbitrary operator,
$O$ consists of left and right factors, depending on $t+x$ and $t-x$ only:

$$O(x,t) = O_L(t+x)\bar O_R(t-x) \eqno(13)$$

\noindent Using the left-right identification we obtain (at $t=0$):

$$ O(x) \to O_L(x)\bar O_L(-x) \eqno(14)$$

\noindent ie. the local operator $O$ becomes effectively bilocal in the
presence of a boundary.  This is similar to the method of image charges in
electrostatics.  An immediate consequence of this is that operators pick up
non-vanishing one-point functions near the boundary (even if they vanish in
the bulk).  ie.

$$<O(x)> \to <O_L(x)\bar O_L(-x)>= {C\over (2ix)^\eta}\eqno(15)$$

\noindent ie. a one-point function becomes a two-point function.  Similarly
two-point functions becomes four-point functions.  In general we can
characterize the effects of the boundary by the operator product expansion:

$$O(x) \to O_L(x)\bar O_L(-x) \to \sum_j{C_j\over
(2ix)^{\eta_j}}O_j(0)\eqno(16)$$

\noindent The set of operators, $O_j$ and exponents, $\eta_j$ simply
correspond to the left-moving Hilbert Space of bulk operators; they do not
depend on the particular boundary conditions.  The operator product expansion
coefficients, $C_j$ on the other hand, do depend on the boundary condition.

How do we find all possible conformally invariant boundary conditions and
calculate Green's functions?  Consider the system in a box of length $l$ with
arbitrary conformally invariant boundary conditions $A$ and $B$ at $x=0$ and
$x=l$, at inverse temperature $\beta$.  The path integral is defined on a
cylinder of circumference $\beta$ and length $l$.  (See Figure 5.)  Letting
$H_{AB}^l$ denote the Hamiltonian for the finite system with boundary
conditions $A$ and $B$; the path integral corresponds to the partition
function:

$$Z_{AB} = tre^{-\beta H^l_{AB}} \eqno(17)$$

\noindent Alternatively, we may regard $l$ as the time interval and $\beta$
as the space interval, ie. make a modular transformation.  The system is now
periodic in space; we write the corresponding Hamiltonian as $H_P^\beta$, $P$
denoting periodic.  Now the system propogates for a time interval $l$ so the
imaginary time evolution operator, $e^{-lH_P^\beta}$ occurs.  However, we do
not write a trace in this case since the system is not periodic in time.
Instead, the system evolves between some initial and final states, $|A>$ and
$|B>$; ie.:

$$Z_{AB} = <A|e^{-lH_P^\beta}|B>\eqno(18)$$

\noindent $|A>$ and $|B>$ are called {\it boundary states}.  To each
boundary condition corresponds a boundary state.  It turns out to be easier
to find all possible boundary states than boundary conditions.  Equating the
two expressions for $Z_{AB}$ gives an equation (true for all $l/\beta$)
which determines all conformally invariant boundary states (and hence
boundary conditions).   The boundary states determine the Greens functions.
(Matrix elements of operators between the boundary state and the vacuum give
the needed operator product expansion coefficients.)  To solve for the
critical behavior of an arbitrary quantum impurity problem we ``just'' have
to find (guess?) the corresponding boundary state.  Frequently the number of
possibilities consistent with the symmetries of a given bulk critical system
is very small.  We have found  boundary states for the various problems
listed in the introduction.

\vspace{10mm} \noindent {\bf 4. Multi-Channel Kondo Effect} \vspace{5mm}

\noindent We generalize the Kondo Hamiltonian of Eq. (2) to include $k$
channels of electrons interacting with a spin-$s$ impurity:

$${\cal H} = \sum_{j=1}^\infty \left[\psi^{\dagger j}{i d\over dx}\psi_j +
\lambda \delta (x)\psi^{\dagger j}{\vec \sigma \over 2}\psi_j\cdot \vec S
\right]\eqno(19) $$

\noindent Note that we have made the (in general unrealistic) assumption that
each channel has identical Fermi velocity (set to $1$), identical Kondo
coupling and identical potential scattering term (assumed to be $0$).  Thus
the model has an $SU(k)$ symmetry corresponding to interchanging the
channels.  When the number of channels exceeds twice the impurity spin, $k >
2s$, the impurity is overscreened and the Kondo coupling does not renormalize
to $\infty$.  Consider, for example the simplest overscreened case, $k=2$,
$s=1/2$. See Figure 6.  If we assume that the on-site Kondo coupling goes to
infinity, then one electron from each channel would sit on the impurity site
with spins anti-parallel to that of the  impurity.  This overscreened complex
has spin $1/2$.  Now consider the electrons on the next site. More correctly
we must consider an even parity combination of the two neighboring sites.  In
the three-dimensional setting we are considering spherical shells around the
impurity, as drawn in Figure 6.  These electrons feel a weak {\it
antiferromagnetic} Kondo coupling with the spin complex at the origin.  This
coupling also renormalizes to large values.  If we assume it also goes to
infinity then we obtain yet another, larger, spin complex with $s=1/2$.  This
process would keep on going forever.  This basically tells us that the
assumption that the Kondo coupling flows to $\infty$ is not correct.

We have found the boundary states for all values of $k$ and $s$.  {}From
these we can calculate the single-particle Green's function [6,\ 9]:

$$<\psi_L^\dagger (z_1)\psi_R(\bar z_2)> \to {S_{(1)}\over z_1-\bar z_2}
\eqno(20)$$

\noindent $S_{(1)}$ measures the correlation of incoming and outgoing
electron.  {}From it we can determine the self-energy for a dilute random
array of impurities, and hence the zero-temperature resistivity.  This is
given by:

$$\rho = \rho_{un}{1-S_{(1)}\over 2} \eqno(21)$$

\noindent Here $\rho_{un}$ is the resitivity in the unitary limit.  ie., it
is the maximum possible resistivity that could occur for potential
scattering, corresponding to a $\pi /2$ phase shift.  $S_{(1)}$ has the
interpretation of the 1 particle $\to$ 1 particle S-matrix at zero energy.
For potential scattering, we would have, $S_{(1)} = e^{2i\delta }$.  This is
also true for the one-channel Kondo effect, with $\delta = \pi /2$.  In
fact, with a particle-hole symmetric Hamiltonian as in Eq. (19), $S_{(1)}$
must be real.  However, in the overscreened case, we find $|S_{(1)}|<1$!!
Unitarity then implies that there must be inelastic scattering even at zero
energy, violating the basic assumption of Landau's Fermi liquid theory.
Thus we call this a non Fermi liquid fixed point.  Our explicit calculation
from the boundary state gives:

$$S_{(1)}={\cos [\pi (2s+1)/(2+k)]\over \cos [\pi /(2+k)]}\eqno(22)$$

\noindent We can also calculate the electron pair-operator Green's function
[6,\ 10], for example.  For the case, $k=2$, $s=1/2$, we define the
spin-singlet, flavor-singlet pair operator:

$$ O (z)\equiv \psi_{L,\alpha i}(z)\psi_{R,\beta j}(\bar z)\epsilon^{\alpha
\beta}\epsilon^{ij}\eqno(23)$$

\noindent  We obtain the Green's function:

$$<O(z_1)O^\dagger (z_2)> = {4\eta^{-1/2}\over |z_1-\bar z_2|^2}\left[{1\over
1-\eta}+3\right]\eqno(24)$$

\noindent Here

$$\eta \equiv {4r_1r_2\over |\tau_1-\tau_2|^2+(r_1+r_2)^2}, \ \  z_j\equiv
\tau_j+ir_j, \ \  (r_j>0)\eqno(25)$$

\noindent This universal cross-over function gives the non-interacting result
in the bulk limit:

$$<O(z_1)O^\dagger (z_2)>\to {4\over |z_1-z_2|^2}\ \  (r_1,
r_2>>|z_1-z_2|)\eqno(26)$$

\noindent and gives non-trivial singular behavior in the boundary limit:

$$<O(z_1)O^\dagger z_2)>\to {8\over |\tau_1-\tau_2|\sqrt{r_1r_2}} \ \
(a<<r_1,r_2 <<|\tau_1-\tau_2|)\eqno(27)$$

\noindent (Here $a$ is the width of the boundary layer, or short-distance cut
off.) Note that the $1/\tau$ fall off is more singular than the $1/\tau^2$
behavior in the non-interacting case. Fourier-transforming with respect to
time for fixed $r_i$ we obtain a logarithmically divergent local pair
susceptibility, $\chi (\omega ) \propto \ln \omega $.  This has been used as
the basis of models of superconductivity [17].

So far, we have calculated the leading temperature dependence of Green's
functions, using only properties of the fixed point.  The flow to the fixed
point is also important and determines various temperature dependent
corrections [2,\ 4,\ 6,\ 9].  These are obtained by considering the leading
irrelevant operator, $O$, at the stable fixed point.  ie., we write the
effective Hamiltonian density:

$${\cal H} = {\cal H}_{FP} + g\delta (x) O\eqno(28)$$

\noindent Here the fixed point Hamiltonian, $H_{FP}$ is defined with the
corresponding boundary condition.  $g$ is the (non-universal) leading
irrelevant coupling constant.  For $k=2$, $s=1/2$, $O$ has scaling dimension
$3/2$.  It then follows from dimensional analysis that $g$ has dimensions of
(energy)$^{-1/2}$.  (Note that the $\delta$-function contributes 1 to the
dimension.)  The corresponding energy scale is, by definition, the Kondo
temperature:

$$ g \equiv {1\over \sqrt{T_K}}\eqno(29)$$

\noindent  We perform low order perturbation theory in $g$.  This is
infrared finite, since $g$ is irrelevant. It gives a series in
$(T/T_K)^{1/2}$, by dimensional analysis.  We can express the leading
temperature dependence of several quantities in terms of one unknown
parameter, 
ratio, exactly for small T.  For $k=2$, $s=1/2$, we find the resistivity
correction [6,\ 9]:

$$\rho (T) = {1\over 2}\rho_{un}[1+4g\sqrt{\pi T}]\eqno(30)$$

\noindent the impurity specific heat [4]:

$$C_{imp}(T) = 9\pi^3Tg^2ln(T/T_K)\eqno(31)$$

\noindent and the impurity susceptibility [4]:

$$\chi_{imp}(T) = 18\pi\mu_B^2g^2\ln (T_K/T)\eqno(32)$$

\noindent Notice that $g = 1/\sqrt{T_K}$, drops out of the Wilson ratio,
$\chi_{imp} /C_{imp}$ and also the ratio of the square of the finite
temperature restivity correction  to the specific heat.   The analogous
ratio for the Fermi liquid Kondo fixed point was first calculated by
Nozi\`eres [15].  In that case the approach to the fixed point is quite
different.  In particular, the resistivity then takes the form:

$$\rho_{FL}(T) = \rho_{un}[1-g^2T^2]\eqno(33)$$

Differences occur both in the dimension of the leading irrelevant operator
(2 instead of 3/2) and the lowest order in perturbation theory at which
various quantities become non-zero.  Note the much more singular T-dependence
in this non-Fermi liquid case.

\vspace{10mm} \noindent {\bf 5. Impurities in Spin-1/2 Heisenberg Chains: A
Fusion Rule Approach} \vspace{5mm}

Recently, Sebastian Eggert and I applied the boundary conformal field theory
technique to a localized impurity in an s=1/2  antiferromagnetic chain
[11].  The impurities we considered were modified exchange couplings on a
single link or a pair of links or the coupling of an extra spin (possibly
with $s>1/2$) to the chain.  The boundary fixed points that occured were
very simple, corresponding to a periodic (ie. unperturbed) chain or an open
chain.  Unlike in our analysis of the multi-channel or two-impurity Kondo
problems [3,\ 4], we did not use the fusion rules to find the fixed point to
which the system renormalized upon adding the impurity.  In this section, I
wish to show that, in fact, it is possible to pass between these fixed
points using fusion, at least in the case of an $SU(2)$ invariant system,
ie. the Heisenberg model.  We do not know how to do this in the case of the
$xxz$ model; it appears likely that fusion is not a general procedure that
works in all cases [13,\ 14].  We hope that this approach to the Heisenberg
model may shed light on this general question.  It also brings out analogies
between the spin chain impurity problem and both two-channel [3,\ 4] and
two-impurity [7] Kondo problems.  We begin by briefly reviewing the fusion
rules approach to finding boundary states and then turn to the spin chain
example.

As mentioned in Sec. 3, the quantum impurity problems that we consider can
always be formulated so as to obey the boundary condition $T(t,0)=\bar
T(t,0)$.  In many cases they also obey an analogous condition on some
conserved chiral currents:

$$ J^a(t,0)=\bar J^a(t,0)\eqno(34)$$

\noindent In the spin chain problem $a$ will run over the three generators of
$SU(2)$. As before, we may regard $\bar J^a$ as the analytic continuation of
$J^a$ to $x<0$.  If we consider the cylinder geometry of Sec. 3, with such
boundary conditions at each end, then it follows that

$$ J^a(t,-l)=J^a(t,l)\eqno(35)$$

\noindent and the same for $T$.  Thus all eigenstates of $H^l_{AB}$ will be
eigenstates of $T$ defined periodically on an interval of length $2l$.  It
follows that $Z_{AB}$ can be expanded in Kac-Moody characters of $T$,
corresponding to the algebra of the currents, $J^a$. These can be written as:

$$\chi_i(q)\equiv \sum_{x;i} q^x\eqno(36)$$

\noindent where the $x$ label the scaling dimensions in the $i^{th}$
conformal tower, and, $q\equiv e^{-\pi \beta /l}$.  ie., in general:

$$Z_{AB} = \sum_in^i_{AB}\chi_i(q).\eqno(37)$$

\noindent Only the integer multiplicities, $n^i_{AB}$, depend on the
particular conformally invariant boundary conditions $A,B$.

Eqs. (11) and (35) correspond to conditions on the boundary states, $A>$,
after the modular transformation:

$$[T(x)-\bar T(x)]|A>=[J^a(x)+\bar J^a(x)]|A>=0\eqno(38)$$

\noindent There is one solution of these equations, called an Ishibashi state
[18], for each Kac-Moody conformal tower:

$$|i> = \sum_N|i,N>_L\otimes \Omega |i,N>_R.\eqno(39)$$

\noindent Here $N$ is summed over all members of the conformal tower; the
anti-unitary operator $\Omega$ takes $\bar J^a \to -\bar J^a$.  All boundary
states, $|A>$ can be expanded in Ishibashi states:

$$|A> = \sum_i|i><i,0|A>\eqno(40)$$

\noindent where $|i,0>\equiv |i,0>_L\otimes \Omega |i,0>_R$.  Thus, the
apparently formidable problem of classifying all boundary states and spectra
is reduced to finding a finite number of multiplicites, $n^i_{AB}$ or a
finite number of matrix elements $<i,0|A>$, one for each conformal tower.  By
equating the two expressions for $Z_{AB}$, Eq. (17) and (18) we obtain an
important set of equations which can be used to determined these parameters.
Using the modular transformation:

$$\chi_i(q) = S^j_i \chi_j(\tilde q),\eqno(41)$$

\noindent where

$$\tilde q \equiv e^{-4\pi l/\beta},\eqno(42)$$

\noindent and assuming linear independence of the characters, we obtain:

$$\sum_jS^i_jn^j_{AB} = <A|i,0><i,0|B>,\eqno(43)$$

\noindent Cardy's equation [1].

Cardy discovered an important property of this equation which allows new
boundary states to be constructed from old ones.  Given any consistent
boundary state, $|A>$, and any conformal tower, $i$, we construct a new
boundary state, $|A,i>$ as follows. With some fixed boundary state $|B>$ at
the other end of the cylinder, the new spectrum is given by:

$$n^j_{A,i;B}=N^j_{ik}n^k_{AB}.\eqno(44)$$

\noindent Here the non-negative integers, $N^j_{ik}$, are  the fusion rule
coefficients; they give the number of occurrances of the primary field
$\phi_j$ in the operator product expansion of $\phi_i$ with $\phi_k$.

\noindent The new boundary state is given by:

$$<A,i|j,0> = {S^j_i\over S^j_0}<A|j,0>.\eqno(45)$$

\noindent The consistency of Eqs. (44) and (45) follows from the Verlinde
formula relating the fusion rule coefficients to the modular S-matrix:

$$S^j_kN^k_{il} ={S^j_iS^j_l\over S^j_0}.\eqno(46)$$

\noindent This strategy for constructing new boundary states was used in our
solution for the critical behavior of the multi-channel and two-impurity
Kondo problems.  We began with a trivial boundary state corresponding to a
simple vanishing boundary condition on the free fermions.  We then
constructed non-trivial boundary states by fusion with particular operators,
$\phi_i$, motivated by physical considerations.

We now consider the example of the $s=1/2$ Heisenberg spin chain.  Our
technique can be applied to this problem because, in the absence of a
boundary, the system corresponds to a harmonic Luttinger liquid, ie. a free
spin boson, or correspondingly a $k=1$, $SU(2)$ Wess-Zumino-Witten (WZW)
non-linear $\sigma$-model.  (See, for example, [19].)  The spin operators,
in the bulk system are represented in terms of the $SU(2)$ matrix field,
$g$, of the WZW model and the current operators as:

$$\vec S_i = (-1)^i\hbox{constant}\times trg\vec \sigma +(\vec J_L+\vec
J_R).\eqno(47)$$

\noindent The field $g$ may be regarded as a product of left and right-moving
factors:

$$g^\alpha_\beta = g_L^\alpha g_{R \beta}\eqno(48)$$

\noindent Impurity interactions generally involve the field $g$ at the
origin.  A crucial complication of this Luttinger liquid system, compared to
free fermions, is that the impurity interaction involves both left and
right-moving fields and cannot be reduced to a single type of field.  This is
unlike the Kondo problem.  If we consider, for example, the Kondo problem in
a one-dimensional system of left and right-moving electrons, the Kondo
interaction only involves the sum of left and right-moving fermion fields,
$[\psi_L(0)+\psi_R(0)]$.  We may define even and odd parity channels,

$$\psi_{e,o}(x)\equiv {1\over \sqrt{2}}[\psi_L(x)\pm \psi_R(-x)],\eqno(49)$$

\noindent and only $\psi_e(0)$ appears in the Kondo interaction.  No analgous
reduction of the number of degrees of freedom can be made in the Heisenberg
model case so we are faced with what is fundamentally a two-channel problem.
It is actually convenient to regard the right-movers as being a second
channel of left movers.  ie., we define:

$$ g_{L\alpha 1}(x) \equiv  g_{L \alpha}(x), \ \  g_{L \alpha 2}(x) \equiv
g_{R\alpha }(-x)\eqno(50) $$

\noindent We now have two left-moving $k=1$ WZW models, a theory with central
charge $c=2$.  This model is known to be equivalent to a single $k=2$ WZW
model ($c=3/2$) together with an Ising model ($c=1/2$).  This representation
is useful because the total spin current, $\vec J_1+\vec J_2\equiv \vec J$,
is conserved whereas the original left and right-moving currents are not
separately conserved in general, with a boundary present.  Thus only a
finite number of characters and Ishibashi states can occur, corresponding to
direct products of $SU(2)_2$ and Ising conformal towers.  Each theory has
three conformal towers.  Those of the WZW model are labeled by the spin of
the highest weight state, $j=0,1/2,1$.  Those of the Ising model correspond
to the identity operator, the order parameter, and the energy operator.  It
is convenient to also label these by a second index $j_I=0,1/2,1$
respectively because the modular $S$-matrices are the same with this
identification: $$S=\left( \begin{array}{ccc} 1/2&1/\sqrt{2}&1/2 \\
1/\sqrt{2}&0&-1/\sqrt{2}\\ 1/2&-1/\sqrt{2}&1/2\end{array}\right) $$

\noindent Let us now consider a boundary.  It turns out that there are
apparently only two relevant boundary conditions corresponding to a periodic
system (ie. no boundary at all) or else an open system, ie. a break in the
spin chain at one point.  As discussed previously, we can always formulate
the problem in terms of left-movers only on the entire real axis or left and
right-movers on the positive axis.  In this problem, in the first
formulation, we have two channels of left-movers and in the second
formulation we have two channels each of left and right-movers.  The latter
formulation is convenient for studying the cylinder geometry with two
boundary conditions $A$ and $B$ discussed in Sec. 4.  We now wish to
determine the partition function $Z_{AB}$ for $A$ and $B$ either the
periodic or open boundary condition, labelled $P$ and $O$ respectively.
Note that $Z_{PP}$ is the partition function for a periodic spin chain of
length $2l$; $Z_{PO}$ corresponds to a chain of length $2l$ with open
boundary conditions and $Z_{OO}$ corresponds to two decoupled chains with
open boundary conditions, each of length $l$. Although we ignore corrections
of $O(1/l)$ to the energies, we must carefully distinguish the case of even
or odd $l$.  These spectra were worked out in [11] by mapping the boundary
conditions on the spins into boundary conditions on the abelian boson which
is equivalent to the $k=1$ WZW model. Periodic boundary conditions on the
spins map into periodic boundary conditions on the boson and open map into
vanishing boundary conditions.  {}From this we determined the partition
functions, written in terms of $k=1$ WZW characters, which are labelled by
highest weight states, $j=0$ and $j=1/2$.  We label the corresponding
characters, $\chi^{(1)}_j$; the superscript $(1)$ labels the Kac-Moody
central charge $k=1$.  The partition functions depend on whether the chains
have even or odd length.  For a periodic chain of even length $2l$ the
partition function is given by
$[\chi^{(1)}_0(q)]^2+[\chi^{(1)}_{1/2}(q)]^2$.  For odd length $2l+1$ it is
given by $2\chi^{(1)}_0(q)\chi^{(1)}_{1/2}(q)$.  For an open chain of even
length, $l$ it is given by $\chi^{(1)}_0(q)$ and for odd length, $l+1$ it is
given by $\chi^{(1)}_{1/2}(q)$.

We can reformulate these spectra in terms of the $SU(2)_2\times$ Ising
representation.  We now give the spectra for the three possible pairs of
boundary conditions ($P$ or $O$ at each end) and for even or odd length of
the
 line segments (denoted by a superscript $e$ or $o$.).  These are given by:

\begin{eqnarray}Z_{PP}^e(q) &=&[\chi^{(2)}_0(q)+\chi^{(2)}_1(q)]\cdot
[\chi^I_0(q) + \chi^I_1(q)] \cr Z_{PP}^o(q) &=&2\chi^{(2)}_{1/2}(q)\cdot
\chi^I_{1/2}(q)\cr Z_{PO}^e(q) &=&[\chi^{(2)}_0(q)+\chi^{(2)}_1(q)]\cdot
\chi^I_{1/2}(q) \cr Z_{PO}^o(q) &=&\chi^{(2)}_{1/2}(q)\cdot [\chi^I_0(q)+
\chi^I_1(q)]\cr Z_{OO}^{ee}(q) &=&\chi^{(2)}_0(q)\cdot
\chi^I_0(q)+\chi^{(2)}_1(q)\cdot \chi^I_1(q)\cr Z_{OO}^{eo}(q)
&=&\chi^{(2)}_{1/2}(q)\cdot \chi^I_{1/2}(q)\cr Z_{OO}^{oo}(q)
&=&\chi^{(2)}_0(q)\cdot \chi^I_1(q)+\chi^{(2)}_1(q)\cdot
\chi^I_0(q)\hskip7.2cm (51)\nonumber\end{eqnarray}

It is straightforward to check all these identities explicitly since the
$SU(2)_2$ and Ising characters all have simple expressions in terms of
$\theta$-functions and the $SU(2)_1$ characters are given by simple free
boson formulas.

We now wish to demonstrate that we can pass  between the different spectra
by fusion.  The fusion rules are the same for $SU(2)_2$ and for the Ising
model using the $j_I$ representation.  They are:

\begin{eqnarray}0\times j &\to j \cr 1/2 \times 1/2 &\to 0 + 1 \cr 1\times
1/2 &\to 1/2 \hskip12cm (52)\nonumber \end{eqnarray}

\noindent Let us consider starting with two open sections, both of even
length, corresponding to $Z_{OO}^{ee}$.  Suppose we now couple one extra
$s=1/2$ variable infinitesimally with equal  strength to the two open chains
at $x=0$. We expect this weak link to ``heal''; ie. to renormalize to a
periodic  boundary condition.  We should then obtain the partition function,
$Z_{PO}^{o}$. (The total length is now odd since we have added one extra
spin.) It is natural to suspect that this process should correspond to
fusion with the $j=1/2$ primary field in the $SU(2)_2$ sector. It can be
verified explicitly from Eqs. (51) and (52) that this is the case.  If we
now couple the remaining open ends to another $s=1/2$ variable then we
obtain the $Z_{PP}^e$ spectrum.  This again can be obtained by fusion with
the $j=1/2$ primary field. Alternatively, we may start with two open
sections both of odd length or one odd and one even.  Adding  extra
$s=1/2$'s, coupled with equal strength at both ends, producing cross-over to
$Z_{PO}$ and then $Z_{PP}$ corresponds to fusion with $j=1/2$ in all cases.
This process is formally identical to a spin-only version of the
two-channel, $s=1/2$ Kondo effect [4]. In that problem the non-trivial fixed
point is also obtained by fusion with $j=1/2$ in the $SU(2)_2$ sector.  The
only difference  lies in the presence of charge excitations in the Kondo
problem; but these play a passive role anyway.

Alternatively, we may again start with two open even length sections and
now  couple them together with {\it two} intervening spins.  This is
analogous to the two-impurity Kondo problem; the coupling of the two extra
spins to the ends of the open chains is analogous to the Kondo coupling and
the self-coupling of the two extra spins is analogous to the RKKY coupling.
(The two Kondo couplings are assumed equal.) The stable fixed points in this
problem  correspond to open chains, with the two extra spins  either
coupling together to form a singlet or else with one fastening onto the end
of each chain.  In between these two stable fixed points we expect an
unstable one, which can be obtained by adjusting the ratio of the
inter-impurity coupling to the Kondo coupling.  This fixed point corresponds
to the periodic boundary condition, with spectrum $Z_{PO}^e$.  It can be
checked from Eq. (51) and (52) that this is obtained from fusion with the
Ising primary, $j_I=1/2$.  Critically coupling the two remaining free ends
to two more impurity spins produces $Z^e_{PP}$, again from fusion with
$j_I=1/2$. Again the same process occurs if we start with two open chains of
odd length or one even and one odd.  This fusion process, with the $j_I=1/2$
field is again identical to the one that determines the non-trivial unstable
critical point in the two-impurity Kondo problem [7].

Rather remarkably, both two channel one impurity and one channel two impurity
processes occur in this simple model, in a simplified form with no charge
degrees of freedom present.  In both cases the ``non-trivial'' fixed point is
simply the periodic one, the ``healing'' process discovered in [11]
corresponds to the flow to the non-trivial fixed point in both types of
Kondo problem, a stable flow for one impurity but unstable for two
impurities.

We may also write down boundary states corresponding to periodic and open
boundary conditions, related to each other by fusion.  Subtleties arise
because of the distinction between even and odd length chains and we find it
neccessary to introduce (at least) two types of periodic boundary states and
three types of open ones, in order to obtain all seven partition functions.
Ishibashi states can be labelled either in terms of the two $SU(2)_1$
algebras or in terms of $SU(2)_2 \times$ Ising.  In the latter formulation
($|j j_I>$ basis), the boundary states are:

\begin{eqnarray}|per\ 1>&=&|0,0>-|0,1>+|1,0>-|1,1>\cr |per\
2>&=&|0,0>+|0,1>-|1,0>-|1,1>\cr |open\ 1>&=&{1\over
\sqrt{2}}[|0,0>+|0,1>+|1,0>+|1,1>+|1/2,1/2>_1+|1/2,1/2>_2]\cr |open\
2>&=&{1\over \sqrt{2}}[|0,0>+|0,1>+|1,0>+|1,1>-|1/2,1/2>_1-|1/2,1/2>_2]\cr
|open\ 3>&=&{1\over
\sqrt{2}}[|0,0>-|0,1>-|1,0>+|1,1>+|1/2,1/2>_1-|1/2,1/2>_2]
\ (53)\nonumber \end{eqnarray}
\noindent Two different Ishibashi states $|1/2,1/2>_i$, $i=1,2$ occur in this
formulation, corresponding to the $SU(2)_1\otimes SU(2)_1$ Ishibashi states
$|1/2,0>$ and $|0,1/2>$.  Using the ratio of $S$-matrix elements:

$${S^j_{1/2}\over S^j_0}=(\sqrt{2},0,-\sqrt{2}),\eqno(54)$$

\noindent we see from Eq. (45) that each of the $|open\ i>$ states goes into
one of the $|per\ j>$ states under fusion with either the $j=1/2$ or
$j_I=1/2$ field.  The 15  matrix elements of the form
$<A|e^{-lH^\beta_P}|B>$ obtained from these five states are all equal to one
of the seven different partition functions given in Eq. (51).  For instance,
$Z_{per\ i,\  per\ j}= Z_{PP}^e$ for $i=j$, and $=Z_{PP}^o$ for $i\neq j$.

To conclude this section, the simple problem of an impurity in a Heisenberg
spin chain can be understood using the fusion rules.  It represents a
spin-only version of both the two channel and two impurity Kondo problems.

I would like to thank Andreas Ludwig for his collaboration in much of this
work and Eugene Wong for checking the identities in Eq. (51). I would also
like to thank my other collaborators in this field: Sebastian Eggert, Jacob
Sagi and Erik S\o rensen. This research was supported in part by NSERC of
Canada.

\vspace{10mm} \noindent {\bf Figure Captions}  \vspace{5mm}

\noindent 1.  Imaginary time formulation of quantum impurity problems.

\noindent 2. Two possible formulation of quantum impurity problems with left
and right-movers on the positive real axis or with left movers only on the
entire axis.

\noindent 3. Renormalization group flow of the Kondo coupling.

\noindent 4. Boundary and bulk limit of correlation functions.

\noindent 5. The cylinder geometry: $A$ and $B$ represent conformally
invariant boundary conditions.

\noindent 6. The two-channel, $s=1/2$ Kondo effect.  Successive electron
orbitals each overscreen the impurity always leaving behind an effective
$s=1/2$.

\vspace{10mm} \noindent {\bf References} \\ \begin{itemize} \parsep=0ex
\itemsep=0ex \item[*] To appear in the proceedings of the Taniguchi Symposium
held at Mie Prefecture, Japan, October, 1993. \item[{[1]}] J.L. Cardy, {\it
Infinite Lie Algebras and Conformal Invariance in Condensed Matter and
Particle Physics} (ed. K. Dietz and V. Rittenberg, World Scientific,
Singapore, 1987), 81;  Nucl. Phys. {\bf B324},  581 (1989); D. Lewellen and
J.L. Cardy, Phys. Lett. {\bf B259}, 274 (1991). \item[{[2]}]  I. Affleck,
Nucl.  Phys.  {\bf B336}, 517 (1990). \item[{[3]}]  I. Affleck and A.W.W.
Ludwig, Nucl. Phys. {\bf B352}, 849 (1991). \item[{[4]}] I. Affleck and
A.W.W. Ludwig, Nucl.  Phys. {\bf B360}, 641 (1991).  \item[{[5]}] I. Affleck
and A.W.W. Ludwig, Phys.  Rev.  Lett. {\bf 67}, 161 (1991). \item[{[6]}]
A.W.W. Ludwig and I. Affleck, Phys.  Rev.  Lett., {\bf 67}, 3160 (1991).
\item[{[7]}] I. Affleck and A.W.W. Ludwig, Phys.  Rev.  Lett. {\bf 68}, 1046
(1992). \item[{[8]}]  I. Affleck, A.W.W. Ludwig, H.-B. Pang and D.L.  Cox,
Phys.  Rev.  {\bf B45}, 7918 (1992). \item[{[9]}] I. Affleck and A.W.W.
Ludwig, Phys.  Rev. {\bf B48}, 7297 (1993). \item[{[10]}]   A.W.W. Ludwig
and I. Affleck, in preparation. \item[{[11]}] S. Eggert and I. Affleck,
Phys. Rev. {\bf B46}, 10866 (1992). \item[{[12]}] E. S\o rensen, S. Eggert
and I. Affleck, preprint, UBCTP-93-07 (cond-mat@babbage.sissa.it, no.
9306003) to appear in J. Phys. {\bf A}. \item[{[13]}] I. Affleck and J.
Sagi, preprint, UBCTP-93-18 (hep-th@babbage.sissa.it, no. 9311056).
\item[{[14]}] E. Wong and I. Affleck, preprint, UBCTP-93-20
(cond-mat@babbage.sissa.it, no. 931140). \item[{[15]}]  P.  Nozi\`eres, J.
Low Temp. Phys. {\bf 17}, 31 (1974);  P. Nozi\`eres, Proceedings of the
$14^{th}$ Int.'l Conf. on Low Temp. Phys. (ed. M. Krusius and M. Vuorio,
North Holland, Amsterdam, 1974) V.5, p. 339. \item[{[16]}] J. Kondo, Prog.
Theor. Phys. {\bf 32}, 37 (1964). \item[{[17]}] D.L. Cox, Phys.  Rev. Lett.
{\bf 67} 2883 (1987);  V.J. Emery and S. Kivelson, Phys. Rev. {\bf B46},
10812 (1992). \item[{[18]}] N. Ishibashi, Mod. Phys. Lett. {\bf A4}, 251
(1989); T. Onogi and N. Ishibashi, Nucl. Phys. {\bf B318}, 239 (1989).
\item[{[19]}] I. Affleck, {\it Fields, Strings and Critical Phenomena} (ed.
E. Br\'ezin and J. Zinn-Justin, North-Holland, Amsterdam), 563 (1990).
\end{itemize} \end{document}